\documentclass{elsart3p}
\usepackage{graphicx}
\usepackage{amsmath}
\usepackage{amssymb}

\begin{document}

\begin{frontmatter}

\title{F\"{o}rster signatures and qubits in optically driven quantum dot molecules}

\author{Juan E. Rolon\thanksref{thank1}}
and
\author{Sergio E. Ulloa}

\address{Department of Physics and Astronomy and Nanoscale and Quantum Phenomena Institute, Ohio University, Athens, OH 45701}

\thanks[thank1]{E-mail: rolon@helios.phy.ohiou.edu}

\begin{abstract}
An interesting approach to achieve quantum gate operations in a solid
state device is to implement an optically driven quantum gate using
two vertically coupled self-assembled quantum dots, a {\em quantum dot molecule} (QDM).
We present a realistic model for exciton dynamics in InGaAs/GaAs QDMs under intense
laser excitation and applied electric fields. The dynamics is obtained by solutions of
the Lindblad master equation. A map of the dressed ground state as function of laser energy and applied
electric field exhibits rich structure that includes excitonic anticrossings that
permit the identification of the relevant couplings. The optical signatures
of the dipole-dipole F\"{o}rster energy transfer mechanism show as splittings of several (spatially) indirect excitonic lines. Moreover, we construct a model for exciton qubit rotations by adiabatic electric field cyclic sweeps into a F\"{o}rster-tunneling regime which induces level anticrossings.  The proposed qubit exhibits Rabi oscillations among two well-defined exciton pairs as function of the residence time at the anticrossing.

\end{abstract}


\end{frontmatter}


An interesting proposal for solid state quantum gate realization is a quantum dot molecule (QDM), a device that may be fabricated by vertically coupled pairs of self-assembled quantum dots \cite{Bayer}. QDMs are promising candidates for quantum information devices due to their ability to allow coherent control of spin and charge degrees of freedom of optically pumped excitons within the constituent dots \cite{Stinaff}.

In this paper we model the exciton dynamics and investigate the optical signatures of quantum dot molecules consisting of two InGaAs/GaAs quantum dots under strong laser excitation and applied electric fields.  We devote special attention to the optical signatures arising from the interplay of interdot tunneling and interdot dipole-dipole interactions that give origin to the F\"{o}rster energy transfer mechanism \cite{Forster}. We incorporate realistic Coulomb energies that already include effects consistent with the experimental finding that strain effects break the symmetry of the two dots \cite{Bester}.

The Hamiltonian of the system is projected in the excitonic basis constructed from direct products of single particle electron and hole states, $\vert e_{T(B)}\rangle$,$\vert h_{T(B)}\rangle$. The subscripts $T$ and $B$ indicate single particles localized at the top and bottom dot respectively and $\vert 0\rangle$ indicates the vacuum. The Hamiltonian can be simplified in the rotating wave approximation, and by performing a unitary transformation that removes the fast oscillating matrix elements \cite{Villas}. We consider exciton states with up to four particles, including single excitons and biexcitons. In the case of a large biexciton detuning, the Hamiltonian can be written in terms of
$\Omega_{T(B)}(t)= \langle \vec{\mu}_{T(B)} \cdot \vec{E}(t)\rangle$, where $\vec{\mu}_{T(B)}$ are the electric dipole moments for the top (bottom) dots, describing the coupling of {\em direct} excitons transitions to the light electric field $\vec{E}(t)$, whereas $t_{h}$ and $t_{e}$ describe the hole and electron tunnelings.  Therefore,
\begin{equation}
H = \left(
\begin{array}{ccccc}
\delta_{0} & \Omega_{T} & 0 & 0 & \Omega_{B} \\
\Omega_{T}& \delta_{eh}^{TT} & t_{e} & t_{h} & V_{F} \\
0 & t_{e} & \delta_{eh}^{BT} & 0 & t_{h} \\
0 & t_{h} & 0 & \delta_{eh}^{TB} & t_{e}\\
\Omega_{B} & V_{F} & t_{h} & t_{e} & \delta_{eh}^{BB}
\end{array}
\right) 
\end{equation}
 The Hamiltonian diagonal elements contain the detuning of the exciton energies with respect to the pumping laser energy; in particular for excitons with indirect character,
\begin{equation}
\delta_{eh}^{BT} = E_{e}^{B}-E_{h}^{T}+U_{eh}^{BT}+\Delta E_{s}-\hbar\omega \nonumber
\end{equation}
\begin{equation}
\delta_{eh}^{TB} = E_{e}^{T}-E_{h}^{B}+U_{eh}^{TB}-\Delta E_{s}-\hbar\omega \, ,
\end{equation}
where we have included the Stark shift $\Delta E_{s}=pF$, for a given applied electric field $F$ and exciton dipole moment $p\sim ed$, $d$ being the interdot distance (in what follows, we use $d=6nm$ ).  We assume the direct exciton confined Stark shift to be vanishingly small, so that only Stark shifts on the indirect excitons are visible \cite{Krenner}.

For typical self-assembled quantum dots, the exciton transition dipole moments $\vec{\mu}_{T}$ and $\vec{\mu}_{B}$ are parallel to each other, normal to the growth direction, and found theoretically and experimentally to be $\mu_{T(B)}\sim 30 Debye$ ($=6.2$ e\AA) \cite{Silverman}.  As a consequence of these values and geometry, strong exciton-radiation coupling and dipole-dipole interaction may take place for direct excitons.  The latter gives origin to the F\"{o}rster resonant energy transfer mechanism, in which a donor quantum dot transfers its exciton energy to the neighboring dot, allowing for non-radiative interdot exciton hoping \cite{ANazir}. The F\"{o}rster coupling $V_{F}$ takes then the form
\begin{equation}
V_F=-\frac{e^2}{4\pi\epsilon_0\epsilon_{r}d^{3}}O_{TB} \, \vec{\mu}_T\cdot\vec{\mu}_B \, ,
\end{equation}
where $\epsilon_r$ is the dielectric constant of the QD and host material.
Assuming unity interdot overlap integrals between T and B excitons, $O_{TB}\sim 1$, and $d=6nm$, one obtains $V_F\sim 2.5meV$, while the interdot tunnelings assume values of $t_h = 1meV$ and $t_e=14.5meV$, respectively.  The similarity of these values suggests one can look for an experimentally detectable signature of the F\"{o}rster interaction.

To obtain the dynamics of the full system including decoherence processes we use the Lindblad master equation \cite{QuantumOpt2}
\begin{equation}
\frac{d\hat{\rho}}{dt}=-\frac{i}{\hbar}[\hat{H},\hat{\rho}]+\hat{L}_{relax}(\hat{\rho}) \, ,
\end{equation}
where the operator $\hat{L}_{relax}(\hat{\rho})$ describes the dissipative part, that for now only includes spontaneous exciton recombination. With the solution of (4) we constructed a map of the Hamiltonian ``dressed" ground state occupation as a function of laser excitation energy and applied electric field, see Fig.\ 1.
\begin{figure}[h]
\includegraphics[width=80mm]{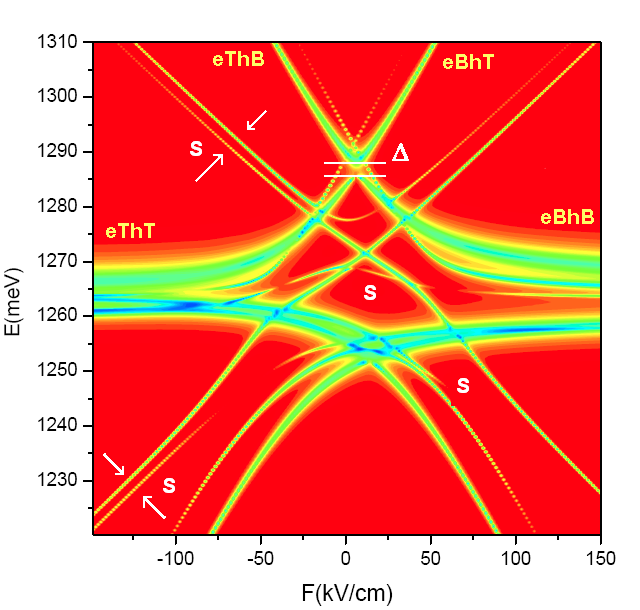}
\caption{Ground state occupation map including biexcitonic signatures. Red color indicates maximum occupation, while green-blue color, minimum occupation. The arrows at S indicate the indirect biexcitonic splitting due to the F\"{o}rster energy transfer coupling $V_F$, while $\Delta=2.5meV$ indicates splitting at the main anticrossing between excitons $\vert e_{B}h_{B}\rangle$ and $\vert e_{T}h_{B}\rangle$ and between $\vert e_{T}h_{T}\rangle$ and $\vert e_{B}h_{T}\rangle$. We have assumed an interband dipole moment of 30Debye.}
\end{figure}

We find rich structure in this map that includes excitonic molecular level anticrossings. The molecular hybridization-like behavior comes via first order processes like single particle tunneling and second order processes like the F\"{o}rster interaction, such that the anticrossing behavior in the map shows the mixing of indirect excitons and direct excitons. Interestingly, the F\"{o}rster coupling is found to split several four-particle (spatially) indirect excitonic lines, in addition to the expected splitting of single direct excitons.  This signature, indicated by $\Delta$ in Fig.\ 1, has not been predicted before.  We propose
that a detailed experimental set up could look for this signature, in spite of having possibly a smaller value due to a weaker spectral overlap between the donor/acceptor excitons.  Clearly the effect is larger for larger transition dipole moments.

To investigate the possible realization of a quantum gate, we exploit the fact that it is possible to identify which excitons mix at a given anticrossing. [As the splitting at the main anticrossing does not change appreciably by the presence of the biexciton state, we ignore the latter in the following for numerical simplicity.] We prepared an input state consisting of selectively excited excitons $\vert e_{T}h_{T}\rangle$,$\vert e_{T}h_{B}\rangle$ plus the vacuum state $\vert 00 \rangle$. The dynamics starts at $t=0$, with an empty molecule state, $\vert 00 \rangle$, as the radiation turns on.  We integrate the dynamics until the transients disappear, and define an starting configuration ($t_i >0$), so that the input state is prepared as superposition
\begin{equation}
\vert\Psi_{I}\rangle = c_{00}\vert 00 \rangle + c_{e_{T}h_{B}}\vert e_{T}h_{B} \rangle+c_{e_{B}h_{B}}\vert e_{B}h_{B} \rangle \, .
\end{equation}
 For $t_{a}>t_{i}$, we then turn on an adiabatic ramp of the electric field, $F(t)$ until $t=t_{ac}$, see Fig. 2.  The electric field sweep correspond to a change from $F_0 = -22.5 kV/cm$, far away from the level anticrossing (see Fig.\ 1), to a final value $F(t_{ac}) = 7 kV/cm$  at the anticrossing.
\begin{figure}[h]
\includegraphics[width=70mm]{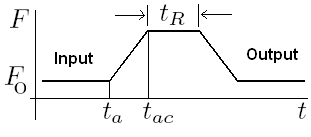}
\caption{Electric field sweep cycle, at fixed laser energy $\hbar\omega=1300meV$, drives the system away from the anticrossing point at $F_{0}=F(t_i)=-22.5kV/cm$ to the anticrossing at $F(t_{ac})=7kV/cm$. The QDM evolves by itself at the anticrossing for a residence time $t_R$. This cycle is repeated for several different $t_R$ values.}
\end{figure}

At $t=t_{ac}$ we fix the electric field (stop the gate pulse) and let the system evolve during a ``residence" time, $t_{R}$. Finally, after $t>t_R+t_{ac}$ we drive the system back to the original configuration (initial electric field) and perform time averages of the resulting output superposition. We repeated the mentioned cycle for several distinct residence times $t_R$ and obtained the average probability of having recovered the different components in $\vert\Psi_{I}\rangle$. We observe clear Rabi oscillations vs $t_R$, predominantly among two pairs of exciton states: ${\vert 00 \rangle, \vert e_{T}h_{B}\rangle}$ (logic 0) and ${\vert e_{T}h_{T}\rangle, \vert e_{B}h_{T}\rangle}$ (logic 1) with a weak background signal given by the optical population of the $\vert e_{B}h_{B}\rangle$ exciton state, thus resembling the evolution of a noisy quasi-two level system,  see Fig. 3.
These oscillations (with period $\simeq 1/\Delta$) signal that one can in principle initialize and rotate {\em at will} the set of excitons states that define suitable qubits.  As the initialization and rotation is done by optoelectronic means, the desirable last step of qubit readout is also in principle possible by monitoring the Stark-shifted luminescence of the system.

\begin{figure}[h]
\includegraphics[width=80mm]{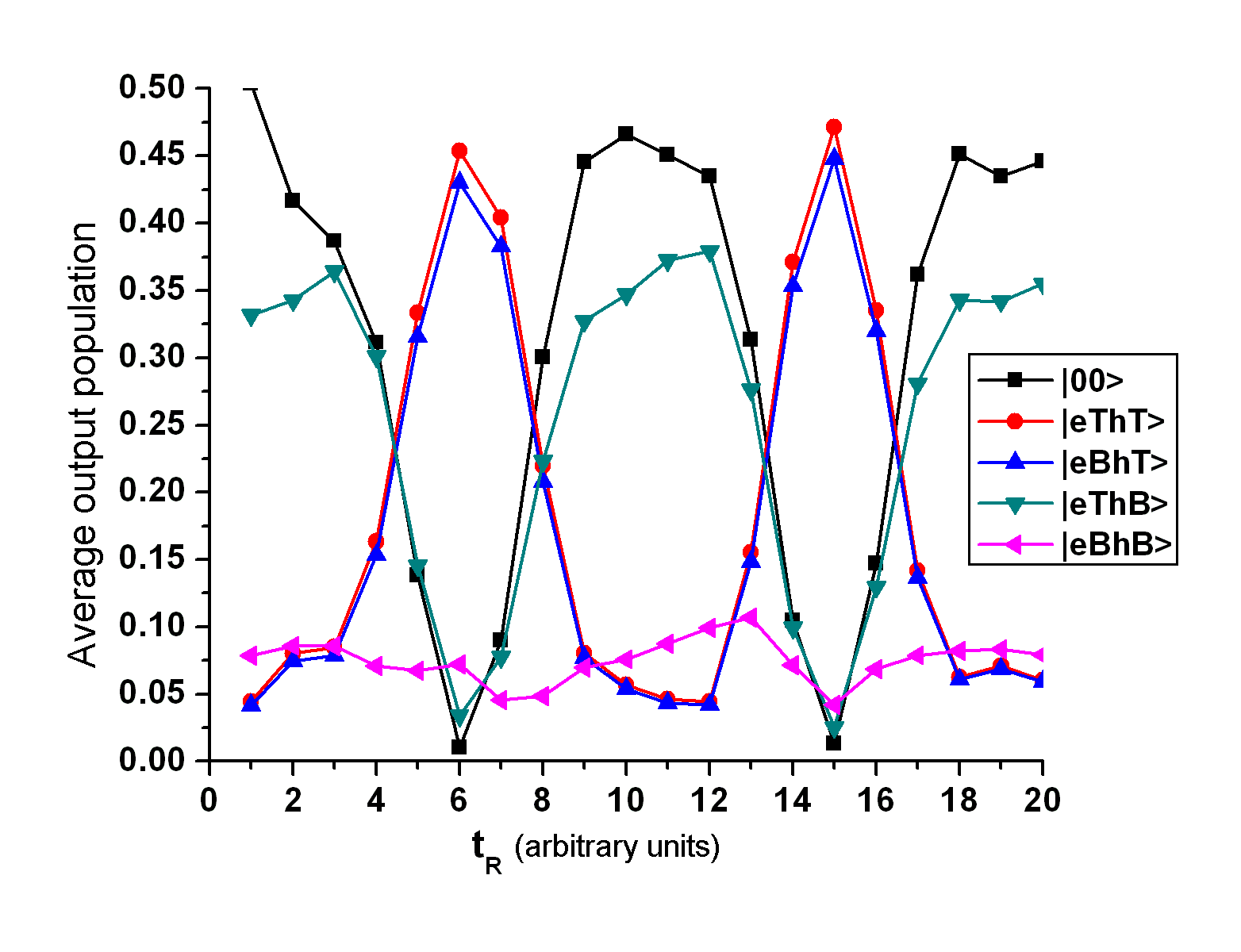}
\vspace{-2.5em}
\caption{Rabi oscillations of two qubit candidates, the pair of excitons, ${\vert 00 \rangle, \vert e_{T}h_{B}\rangle}$ (green,black) and ${\vert e_{T}h_{T}\rangle, \vert e_{B}h_{T}\rangle}$ (blue,red). In the background, the exciton $\vert e_{B}h_{B}\rangle$ (pink) may lead to a noisy manipulation of the desired qubits}
\end{figure}

We have investigated a model for a quantum dot molecule under strong laser illumination and applied electric fields. We have used realistic parameters and have found interesting optical signatures deriving from a competition between single particle tunneling and the exciton F\"{o}rster energy transfer mechanism, with possible means to identify its signature. We have found that with careful sweep cycles of the electric field, it is possible to achieve a basic quantum gate that rotates the proposed qubit consisting of two pairs of exciton states.

\end{document}